\documentclass[10pt, conference]{IEEEtran}
\usepackage{amsmath,amssymb,amsfonts}
\usepackage{pgfplots}
\pgfplotsset{compat=1.18}
\usepackage{algorithmic}
\usepackage{graphicx}
\usepackage{textcomp}
\usepackage{xcolor}
\usepackage{algorithm}

\usepackage{booktabs} 
\usepackage{subcaption}
\DeclareCaptionFont{customsize}{\fontsize{9.5pt}{10.5pt}\selectfont}
\usepackage{fancyvrb}
\usepackage{multirow}
\usepackage{diagbox}
\usepackage{tikz}
\usepackage{numprint}
\usepackage{url}
\usepackage{colortbl}
\usepackage{dhucs} 

\usepackage{enumitem} 
\usepackage{makecell} 
\usepackage[noadjust]{cite} 

\newcommand\refFigure[1]{Fig.~\ref{#1}}
\newcommand\refTable[1]{Table~\ref{#1}}
\newcommand\refSection[1]{Section~\ref{#1}}

\newcommand\korean[1]{}
\newcommand\blind[1]{XXXX}

\newcommand\LKW[1]{\textcolor{black}{#1}}

\setlist[itemize]{leftmargin=*, itemsep=1pt, topsep=2pt}

\usepackage{amsmath}   
\renewcommand{\baselinestretch}{0.97}
\IEEEoverridecommandlockouts
\IEEEaftertitletext{\vspace{-0.5\baselineskip}}     

\begin{document}

\title{\huge CHEQ: A Concurrent Hardware Evaluation Platform for QEC Decoders via Batch-Invariant Circuit-Level Noise Sampling
}
\IEEEaftertitletext{\vspace{-4ex}}

\author{
Sechan Park, Kyeongwon Lee, Mundo Jung, Chaebin Jung, and Woojoo Lee\\
School of Intelligent Semiconductor Engineering, Chung-Ang University, Seoul, Korea \\

\thanks{
This work was supported in part by the National Research Foundation of Korea (NRF) grant funded by the Korea government (MSIT) (No. RS-2024-00345668). Corresponding author: Woojoo Lee

Sechan Park and Kyeongwon Lee Lee contributed equally to this work.


}
}
\maketitle

\begin{abstract}
\boldmath
Hardware quantum error correction (QEC) decoders require validation of logical error suppression and timing compliance before deployment.
Rare logical failures and deadline violations demand extensive sampling, while algorithm-level software evaluation does not directly measure hardware response times.
Circuit-level noise models contain many error mechanisms but few firings per shot at low physical error rates, making exhaustive per-shot scanning inefficient.
We present \textbf{CHEQ}, an FPGA platform integrating on-chip sampling, concurrent decoder evaluation, scoring, and RISC-V control.
Its Sampling Units use geometric skipping to sample firing locations from detector error models and amortize mechanism traversal across shots within each batch.
Under independent random draws and exact geometric sampling, the shot distribution is invariant to batch size, enabling a memory--throughput tradeoff.
A throughput model guides joint selection of the Sampling Unit count, batch size, and decoder count under FPGA resource constraints.
Across evaluated conditions, the pooled one-sided 95\% upper bound on the weighted root-mean-square relative firing-rate difference from Stim is $0.16\%$.
At a batch size of 4,096, a Sampling Unit uses only $4.8\%$--$11.2\%$ of a Helios decoder's LUTs for code distances 7, 9, 11, and 13.
On an AMD VCU118 FPGA at code distance 7 and physical error rate $10^{-3}$, two Sampling Units supply 20 Helios decoders at 49.2 million shots/s, completing $10^{10}$ shots in 3.4 minutes without host-side shot generation.
\unboldmath
\end{abstract}

\section{Introduction}\label{sec:intro}

As quantum computing progresses from the noisy intermediate-scale quantum era toward fault-tolerant operation, quantum error correction (QEC) decoders are becoming essential components of quantum computing systems~\cite{Preskill:Quantum18, Das:HPCA22, Bausch:Nature24, DATE26:Keynotes}.
A QEC decoder is a classical processor that infers errors from repeatedly measured syndromes and determines corrections.
In superconducting systems, syndrome rounds arrive approximately every $1\,\mu\mathrm{s}$, requiring decoding throughput to keep pace~\cite{Google:Nature25, Ziad:NatCommun25, Caune:NatCommun26}.
Sustained throughput deficits cause syndrome backlogs, while long response times can violate time-sensitive feedback deadlines~\cite{Terhal:RMP15, Tan:PRXQ23, Caune:NatCommun26}.
These throughput and latency constraints motivate dedicated FPGA and ASIC implementations~\cite{Ueno:DAC21, Liyanage:TQE24, Barber:NatElectron25, Wu:ASPLOS25}.
For surface codes below threshold, increasing the code distance exponentially suppresses logical errors but also increases the syndrome volume per round and decoding-graph size without extending the syndrome interval.
Although recent superconducting processors have demonstrated improved logical error suppression with increasing distance, practical applications may require logical error rates several orders of magnitude lower than those demonstrated~\cite{Gidney:Quantum21, Beverland:arXiv22, Google:Nature25}.
These requirements continue to motivate faster and more scalable decoder designs~\cite{Overwater:TQE22, Maurya:EuroSys26, Maan:NatCommun26}.

Before deployment, hardware QEC decoders must be evaluated for both logical error suppression and timing compliance~\cite{Battistel:NanoFut23, Maurya:IISWC25, Molavi:arXiv26}.
Algorithm-level software evaluation can characterize decoding decisions but does not directly measure target-hardware response times; evaluating the implementation as a device under test (DUT) exposes both in the same run.
Because logical failures and deadline violations must be rare, direct evaluation requires large numbers of sampled decoding instances, or shots~\cite{GarciaHerrero:Quantum26, Molavi:arXiv26}.
For direct Monte Carlo estimation at fixed relative precision, the required shot count scales inversely with failure probability.
Large-scale applications can require logical error probabilities below $10^{-10}$ per logical operation~\cite{Beverland:arXiv22, Google:Nature25}; for perspective, a per-shot logical failure probability of $10^{-10}$ yields only one expected failure in $10^{10}$ shots.
At this scale, evaluation throughput depends on the complete path from shot generation through DUT delivery to result scoring, rather than generation alone.
External generation incurs per-shot off-chip transfer, while storing pregenerated shots requires space proportional to the shot count.
Generating and consuming shots on the same device removes this external transfer from the per-shot evaluation path.

Recent studies have explored on-chip noise generation for hardware QEC decoder evaluation~\cite{Bascones:EPJQT25, GarciaHerrero:Quantum26}.
Because the generator and DUTs share device resources, greater generation parallelism leaves fewer resources for decoding, whereas insufficient parallelism can make generation the bottleneck, particularly at low physical error rates~\cite{GarciaHerrero:Quantum26}.
Realistic evaluation further requires circuit-level noise, where faults in syndrome-extraction gates and measurements propagate into correlated detector patterns~\cite{Fowler:PRA12, Higgott:PRX23}.
At larger code distances, such models can contain tens of thousands of error mechanisms, although only a small fraction fire in each shot.
Testing every mechanism per shot therefore devotes most sampling work to non-firing events, motivating hardware that samples firing locations and amortizes mechanism traversal across shots.

We present \textbf{CHEQ} (Concurrent Hardware Evaluation platform for QEC decoders), an FPGA platform that generates circuit-level noise samples on chip and concurrently evaluates the logical error suppression and timing compliance of multiple decoder DUTs.
Each Sampling Unit accepts a detector error model (DEM) exported by Stim~\cite{Stim:Quantum21} and uses geometric skipping to sample each mechanism's firing locations within a batch.
One DEM traversal serves an entire batch, amortizing traversal overhead while event-processing work scales with actual firings.
Under independent random draws and exact geometric sampling, the shot distribution is invariant to batch size, enabling a memory--throughput tradeoff.
Shots from each completed batch are distributed among multiple DUTs, allowing one Sampling Unit to feed several DUTs concurrently.
A throughput model guides joint selection of Sampling Unit count, batch size, and DUT count under FPGA resource constraints.
CHEQ integrates Sampling Units, a DUT cluster, a Decoder Interfacing Unit, and a Scoring Unit with a RISC-V processor on one FPGA to execute evaluation campaigns without exchanging bulk shot data with an external host.

We evaluate CHEQ using Helios decoder instances~\cite{Liyanage:TQE24} at physical error rates of $10^{-3}$ and $5\times10^{-4}$.
In the evaluated configurations, each Sampling Unit generates shots faster than a single DUT consumes them, enabling multiple DUTs to share its generation capacity.
Comparisons with Stim using the same DEMs yield a pooled one-sided $95\%$ upper confidence bound of $0.16\%$ on the weighted root-mean-square relative difference in per-mechanism firing rates.
At a batch size of 4,096, a Sampling Unit uses $4.8\%$--$11.2\%$ of a single Helios decoder's LUTs across the evaluated code distances 7--13.
At $d=7$ and $p_{\mathrm{phys}}=10^{-3}$, two Sampling Units sustain 20 DUTs at 49.2 million shots/s on a VCU118.

\section{QEC Decoder Evaluation at Scale}
\label{sec:scale}

\subsection{Evaluation Requirements and Circuit-Level Noise}
\label{sec:prelim}

Direct Monte Carlo estimation of a rare logical failure probability requires a shot count that scales inversely with that probability at a fixed relative precision.
Decoder response times also depend on the input syndrome, so characterizing rare deadline violations requires sufficient samples from the latency tail~\cite{Google:Nature25, Maurya:EuroSys26}.
Logical failures corrupt encoded information, whereas deadline violations can delay dependent logical operations and violate time-sensitive feedback constraints~\cite{Tan:PRXQ23, Caune:NatCommun26}.
Software rare-event methods can reduce the sampling cost of logical failure estimation by modifying the sampling distribution~\cite{Heussen:PRR24, Mayer:arXiv25, Beverland:arXiv25}.
These estimates alone, however, do not directly characterize the response times of the target hardware under the target distribution.
Evaluating the hardware as a DUT with shots drawn from this distribution directly exposes both its decoding decisions and response times.
To avoid generation becoming the evaluation bottleneck, shots must be supplied at least as fast as the DUT consumes them.

Circuit-level evaluation must preserve the detector patterns and correlations induced by the target syndrome-extraction circuit~\cite{Higgott:PRX23}.
In circuits using ancilla qubits, a Pauli fault can propagate through entangling gates and flip detectors at different spatial locations and measurement rounds~\cite{Fowler:PRA12}.
Because both decoding decisions and response times depend on these patterns, the evaluation workload must capture the corresponding circuit-level error statistics.
Relevant fault locations include gates and measurements, as well as idle intervals when specified by the target noise model~\cite{deMartiiOlius:Quantum24}.

A DEM captures these effects and enables shot generation without re-simulating the circuit for every shot~\cite{Stim:Quantum21}.
It represents the noise as a set of independent stochastic error mechanisms, each specifying a firing probability and the detector and logical-observable bits flipped upon firing.
For stabilizer circuits with stochastic Pauli noise, the effect of each fault on these bits can be precomputed by propagating the fault through the circuit~\cite{Derks:Quantum25}.
A shot is then formed by independently sampling the mechanisms and XOR-accumulating the flip targets of those that fire.
Although the mechanisms are independent, detector flips can be correlated because a single mechanism may affect multiple detectors.
Direct sampling from a DEM with $M$ mechanisms therefore replaces per-shot circuit simulation with $M$ Bernoulli draws and XOR accumulation over the firing mechanisms.
For the surface-code circuits considered in this work, the mechanism count scales as $O(d^{2}r)$ with code distance $d$ and the number of syndrome rounds $r$.
In the evaluated configurations, $M$ increases from 5,471 at $d=7$ to 41,909 at $d=13$.

\vspace*{-0.4em} 
\subsection{Batch-Invariant Firing-Based Shot Generation}
\label{sec:sampling}
\vspace*{-0.2em}

Geometric-skip sampling avoids testing non-firing events individually by drawing successive firing positions directly.
If mechanism $i$ fires with probability $p_i$, the expected number of firing mechanisms per shot is $\Lambda=\sum_{i=1}^{M}p_i$, with $\Lambda\ll M$ at low physical error rates.
For a fixed mechanism, firings across shots form independent Bernoulli trials with probability $p_i$, so the number of non-firing shots before the next firing follows a geometric distribution.
Sampling this gap jumps directly to the next firing without testing the intervening shots.
Stim similarly uses geometric skipping for low-probability sampling~\cite{Stim:Quantum21}.

The Sampling Unit traverses the mechanism list sequentially with a shared datapath, avoiding per-mechanism sampling logic that would compete with the decoder DUTs for FPGA resources~\cite{GarciaHerrero:Quantum26}.
Repeating this traversal for every shot would incur an $O(M)$ per-shot cost despite sparse firings, so the Sampling Unit amortizes one traversal over a batch of $W$ shots while using geometric skipping to enumerate each mechanism's firing positions.
Mechanism $i$ contributes $Wp_i$ expected firings plus one terminating draw per batch, yielding $M+W\Lambda$ draws per batch, or $M/W+\Lambda$ per shot.
As $W$ increases, the per-shot draw count approaches $\Lambda$ rather than $M$.

In hardware, $W$ determines the batch-buffer depth and associated control structure and is fixed at design time.
Under independent random draws and exact geometric sampling, the generated shot distribution is invariant to $W$.
By the memoryless property of the geometric distribution, drawing a fresh gap at a batch boundary is distributionally equivalent to continuing from the residual gap of an uninterrupted Bernoulli firing process.
Restarting each mechanism at batch boundaries therefore preserves its firing process in distribution.
Because the mechanisms remain independent and shots are formed by XOR accumulation, batching does not change the joint shot distribution.
Finite-precision deviations and consistency across the evaluated $W$ values are examined in {\refSection{sec:eval}}.

\begin{figure*}[!t]
\vskip -6pt
\centering
\includegraphics[width=\textwidth]{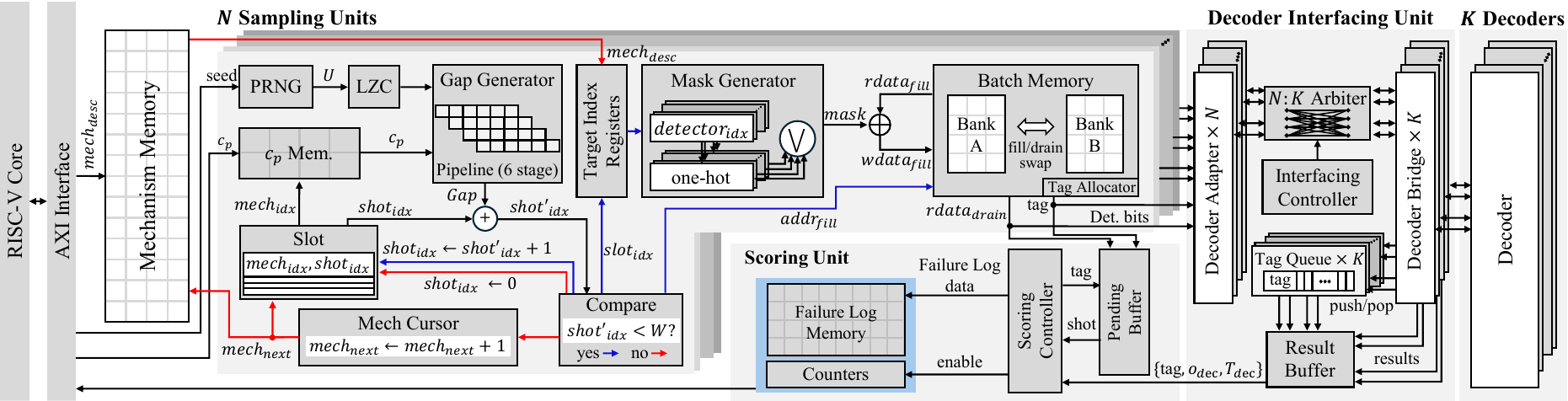}
\vskip -4pt
\caption{Block diagram of the CHEQ platform on a single FPGA.
Shot rows flow left to right from the Sampling Units through the Decoder Interfacing Unit to $K$ decoder DUTs and the Scoring Unit; the RISC-V core loads the configuration beforehand, drains failure logs during the campaign, and reads the counters upon completion.}
\label{fig:arch}
\vskip -12pt
\end{figure*}

Batch size can therefore tune the memory--throughput tradeoff without changing the target shot distribution.
Amortizing the per-mechanism traversal and termination overhead across $W$ shots gives the following first-order model for the average generation time per shot, excluding downstream stalls:
\begingroup
\setlength{\abovedisplayskip}{2pt} 
\setlength{\belowdisplayskip}{2pt} 
\begin{equation}
T(W) = aM/W + b\Lambda ,
\label{eq:tw}
\end{equation}
\endgroup
where $a$ captures the effective fixed cost incurred once per mechanism per batch, including traversal and termination, and $b$ captures the incremental cost associated with an in-batch firing event.
Both coefficients are implementation dependent and absorb the effects of pipeline overlap.
Only the first term decreases with $W$, so increasing the batch size reduces the amortized traversal overhead at the cost of a larger batch buffer, while $T(W)$ approaches the firing-limited term $b\Lambda$.
When $T(W)$ becomes smaller than the per-shot service interval of a decoder DUT, one Sampling Unit can supply multiple DUTs; increasing $W$ enlarges this headroom until the firing-dependent term or FPGA resources become the limiting factor.

\section{CHEQ Evaluation Engine Architecture}
\label{sec:design}
\vspace{-0.1em}

\refFigure{fig:arch} shows the CHEQ evaluation engine integrated into a RISC-V-based FPGA platform.
The engine implements the complete evaluation path from shot generation and distribution to decoding and result scoring on a single FPGA.
It comprises an AXI Interface for processor access, a Mechanism Memory providing one descriptor read port per Sampling Unit, $N$ Sampling Units that generate shots, a Decoder Interfacing Unit (DIU) that dispatches them to $K$ decoder DUTs and collects their responses, and a Scoring Unit that evaluates results.

\subsection{Sampling Unit}
\label{sec:sampling_unit}
\vspace{-0.1em}

Each Sampling Unit generates batches of $W$ shots by drawing firing positions rather than individual mechanism--shot decisions.
DEM preprocessing produces a sampling coefficient $c_p$ and a detector/observable flip descriptor $mech_{\mathrm{desc}}$ for each mechanism, stored in the unit's $c_p$ memory and shared Mechanism Memory, respectively.
The Batch Memory accumulates shot vectors in DEM-defined detector and observable bit order and delivers completed batches to the DIU.

\vspace{0.2em}
\noindent\textbf{Gap generation.}
For firing probability $p$ and a uniform variate $U\in(0,1)$, inverse-transform sampling gives $Gap=\lfloor(-\log_2 U)c_p\rfloor$, where $c_p=-1/\log_2(1-p)$.
Precomputing $c_p$ replaces division with multiplication.
A hardware xoshiro128++ PRNG~\cite{Blackman:TOMS21} supplies one 32-bit value representing $U$ per cycle.
Since $c_p\approx0.693/p$ for small $p$, each coefficient uses an 18-bit normalized significand and a 4-bit shift.
The Gap Generator approximates $-\log_2 U$ using a leading-zero count and a 256-entry $\log_2$ table with linear interpolation, then multiplies by the coefficient significand and restores the scale using the shift.
Its six-stage pipeline sustains one gap per cycle.
A zero random input or a computed gap of at least $W$ is saturated to $W$, placing the next firing outside the batch.
Coefficient quantization, logarithm approximation, and finite random-number precision are evaluated in \refSection{sec:exp_result}.

\vspace{0.1em}
\noindent\textbf{Interleaved scheduling.}
Successive draws of the same mechanism have an eight-cycle recurrence because each depends on the preceding position update.
Eight interleaved Slots hide this latency, each storing $mech_{\mathrm{idx}}$ and the next search position $shot_{\mathrm{idx}}$.
At batch start, the Mech Cursor assigns the first eight mechanisms with zero search positions.
The Slots issue draws in round-robin order using their $c_p$ values and successive PRNG outputs, allowing each update to complete before the Slot is revisited.
For a returned gap, $shot'_{\mathrm{idx}}=shot_{\mathrm{idx}}+Gap$.
If $shot'_{\mathrm{idx}}<W$, the unit emits an event and advances to $shot'_{\mathrm{idx}}+1$; otherwise, the Mech Cursor assigns the next mechanism with a zero search position or empties the Slot if none remain.

\noindent\textbf{Sparse mask generation.}
Each descriptor stores observable flip bits and sparse detector-index entries with valid bits rather than a full-width shot mask.
Four detector entries suffice for the evaluated DEMs, with unused entries marked invalid and repeated indices reduced modulo two during preprocessing.
The Target Index Registers load one descriptor per Slot upon mechanism assignment and reuse it for every firing.
The Mask Generator OR-combines one-hot vectors of the valid detector indices and appends the observable flip bits to form a shot-width $mask$.
XOR-accumulated observable bits provide the reference observable for scoring.

\vspace{0.1em}
\noindent\textbf{Double-buffered accumulation.}
Two Batch Memory banks overlap filling the current batch with draining the preceding one.
Each event reads the row at $shot'_{\mathrm{idx}}$, XORs it with $mask$, and writes it back, making the final shot independent of event order.
The read--modify--write pipeline accepts one event per cycle, matching the maximum arrival rate of one event per draw, while a bypass handles successive accesses to the same row before writeback.
Both banks are cleared at campaign initialization, and drained rows are cleared for reuse.
A fill completes after the mechanism list is exhausted, all Slots are empty, and all writes have committed; the banks exchange roles after both fill and drain complete.
The next batch restarts the Mech Cursor at the beginning of the DEM while retaining the PRNG state to avoid repeating the same sequence.

The model in \refSection{sec:sampling} predicts $M+W\Lambda$ draws per batch: $W\Lambda$ expected firings and one terminating draw per mechanism.
Because descriptor access, gap generation, and event accumulation overlap, the expected fill cost is $C_{\mathrm{fill}}\approx M+W\Lambda$ cycles after neglecting pipeline startup and final Slot retirement, giving the coefficients $a=b=1$ in \eqref{eq:tw}.

\vspace{-0.3em}
\subsection{Decoder Evaluation Path}
\label{sec:dec_eval}
\vspace{-0.2em}

\noindent\textbf{Tagged dispatch.}
Variable decoding latency across $K$ concurrent DUTs can produce out-of-order results, while only detector bits enter the DUT and the reference observable must be retained for scoring.
CHEQ uses the index of a free Pending Buffer entry as each shot's tag and retains the complete shot in that entry until scoring.
Its detector bits and tag are forwarded to the Decoder Adapter, which selects and rearranges the required detectors to match the DUT's input format.
For the unmodified Helios circuit-level union-find decoder~\cite{Liyanage:TQE24}, the Adapter formats the $Z$-subgraph detectors as $r+1$ detector layers of $(d^2-1)/2$ bits.
An $N{:}K$ Arbiter dispatches each adapted input and tag to an idle Decoder Bridge, which transfers the input according to the DUT handshake and retains the tag in its Tag Queue.
Because the Bridges operate independently, other shots can be dispatched while earlier DUTs are receiving or processing their inputs.

\begin{table}[t]
  \vskip -4pt
  \centering
  \caption{Independent deadline and logical scoring.}
  \label{tab:scoring}
  \vskip -4pt
  \setlength{\tabcolsep}{2pt}
  \renewcommand{\arraystretch}{1.1}
  \footnotesize
   \begin{tabular*}{\columnwidth}{@{}c@{\extracolsep{\fill}}ccc@{}}
     \toprule
 Response & Deadline violation & Logical failure & Failure log \\
  \midrule
  On-time match     & --       & --            & -- \\
  ~~On-time mismatch~~   & --       & Counted       & ~~~Recorded~~~ \\
  Late match        & Counted  & --            & ~~~Recorded~~~ \\
  Late mismatch     & Counted  & Counted       & ~~~Recorded~~~ \\
  Timeout           & Counted  & Not evaluated & ~~~Recorded~~~ \\
  \bottomrule
\end{tabular*}
  \vskip -12pt
\end{table}

\vspace{0.1em}
\noindent\textbf{Response tracking and scoring.}
A Bridge counter measures $T_{\mathrm{dec}}$ from acceptance of the final input layer to emission of the final result bit.
Because input layers arrive sequentially in real-time operation, this interval captures the post-input latency relevant to downstream feedback; it includes result output but excludes input loading and pre-dispatch waiting.
The processor sets an evaluation deadline and a larger maximum wait.
If no complete response arrives by the latter, the Bridge resets the DUT and records the limit as a timeout bound rather than a measured completion latency.
For a completed response, $o_{\mathrm{dec}}$ is the parity of the returned correction bits selected by a processor-programmed observable mask.
The Bridge forwards each tagged response or timeout record to the Result Buffer.

The Scoring Controller uses the returned tag to retrieve the original shot and releases its Pending Buffer entry after scoring.
\refTable{tab:scoring} summarizes the policy: every completed response is checked for logical mismatch, and one exceeding the deadline is additionally counted as a deadline violation; a timeout is counted only as a deadline violation.
For each failed shot, the Failure Log Memory stores its batch and shot indices, original shot, available DUT correction data, $T_{\mathrm{dec}}$, and deadline-violation status.



\vspace{0.1em}
\noindent\textbf{DUT integration and throughput.}
DUT-specific interface logic is confined to the Decoder Adapters and Decoder Bridges.
Retargeting requires adapting only these two blocks to the DUT's input and output conventions, while the DEM and decoding graph must correspond to the same syndrome-extraction circuit and detector definitions.
Each Bridge waits for DUT input readiness, allowing non-overlapped DUTs that accept a new shot only after completing the current one to be evaluated without modification.

A full Pending Buffer stalls drain reads and delays bank exchange, so throughput is limited by the slower of generation and downstream consumption.
For each Sampling Unit, excluding bank-switch overhead, the steady-state batch period is $\max(C_{\mathrm{fill}},C_{\mathrm{drain}})$, where $C_{\mathrm{drain}}$ includes delivery and clearing of the $W$ shots and stalls from DUT availability and buffer occupancy.
The average DUT service interval $C_{\mathrm{dec}}$ spans input loading, decoding, and result output, making it distinct from the response latency $T_{\mathrm{dec}}$.
With $T(W)$ also in cycles per shot, \refSection{sec:config} jointly selects $N$, $W$, and $K$ using these costs and the FPGA resource budget.


\vspace{0.1em}
\noindent\textbf{Campaign control.}
Before execution, the RISC-V processor loads $mech_{\mathrm{desc}}$, $c_p$, the observable mask, maximum wait, evaluation deadline, one PRNG seed per Sampling Unit, and the total batch count $B$ through the AXI Interface, and then issues a single start command.
It drains failure logs to DDR during execution and reads the aggregate counters after all $n=BW$ shots are evaluated.
All generation, decoding, and scoring remain on chip; after configuration, only aggregate counters and failure logs cross the FPGA boundary.

\vspace{-0.2em}
\subsection{Configuration Selection}
\label{sec:config}
\vspace{-0.3em}

For a given workload and clock frequency, CHEQ jointly selects $N$, $W$, and $K$ to maximize campaign throughput within the FPGA resource budget.
With sufficient buffering and no DIU or Scoring Unit bottleneck, the steady-state throughput in evaluated shots per cycle is modeled as
\begin{equation}
Q(N,K,W)
=
\min\!\left(
N/T(W),\ K/C_{\mathrm{dec}}
\right),
\label{eq:q}
\end{equation}
\noindent where $T(W)=M/W+\Lambda$ is the per-shot generation cost and $C_{\mathrm{dec}}$ is the average DUT service interval.
For each candidate $N$, $K$ is set to the maximum resource-feasible value because $Q$ is nondecreasing in $K$.
Conversely, once the aggregate sampling rate reaches the aggregate DUT service rate $K/C_{\mathrm{dec}}$, additional sampling capacity does not increase throughput.
Moreover, $T(W)>\Lambda$ for any finite $W$, so the number of DUTs that one Sampling Unit can keep continuously supplied remains below $C_{\mathrm{dec}}/\Lambda$.
Supplying a DUT cluster beyond this limit therefore requires additional Sampling Units rather than batch expansion alone.

For this selection model, the $W$-dependent resource cost is represented by the Batch Memory allocation.
For each feasible $N$, $W_{\max}(N)$ denotes the largest per-unit batch size for which the $N$ double-buffered Batch Memories fit within the URAM budget.
The maximum DUT count supported by the remaining resources, accounting for the Sampling Units, Mechanism Memory, per-DUT interface logic, and platform overhead, is denoted by $K_{\mathrm{res}}(N)$.
The maximum throughput for this $N$ is therefore $Q_N=\min\{N/T(W_{\max}(N)),\,K_{\mathrm{res}}(N)/C_{\mathrm{dec}}\}$.
CHEQ evaluates $Q_N$ over all feasible $N$, chooses the smallest $N$ attaining the maximum predicted throughput as $N^*$, and sets $K^*=K_{\mathrm{res}}(N^*)$.

The batch size is then reduced to the smallest Batch Memory allocation that preserves $Q_{N^*}$.
When the selected configuration is DUT-limited, the condition
$N^*/T(W)\ge K^*/C_{\mathrm{dec}}$
gives
\begin{equation}
W^*
=
\left\lceil
\frac{M}
{N^*C_{\mathrm{dec}}/K^*-\Lambda}
\right\rceil_{D_u},
\label{eq:wstar}
\end{equation}
where $\lceil x\rceil_{D_u}=D_u\lceil x/D_u\rceil$ rounds $x$ up to a multiple of the URAM depth $D_u$, which determines the Batch Memory allocation granularity.
The denominator is positive whenever the DUT-limited rate is achievable with a finite batch size.
If sampling remains the bottleneck, $W^*=W_{\max}(N^*)$ is retained to maximize the predicted throughput.
For the selected $N^*$, the descriptor table is replicated across $\lceil N^*/2\rceil$ identical dual-port memories, providing one dedicated read port to each Sampling Unit.

\vspace{-0.1em}
\section{Experimental Evaluation}
\label{sec:exp_result}
\vspace{-0.2em}

We evaluate the Sampling Unit's fidelity to the target distribution and its ability to generate shots faster than a single DUT consumes them.
We then characterize the campaign throughput, DUT response latency, and logical error rates of configurations selected using the method in \refSection{sec:config}.

\vspace{-0.3em}
\subsection{Evaluation Methodology}
\label{sec:setup}
\vspace{-0.2em}

\noindent\textbf{Implementation.}
The CHEQ evaluation engine is implemented in Verilog RTL and integrated into a stand-alone FPGA system built upon the hardware infrastructure developed for our prior quantum-circuit emulation work~\cite{Choi:TC26,Choi:QIP26,Park:ICCAD26}, comprising a Rocket RISC-V core~\cite{Rocket}, DDR4 controller, AXI interconnect, and peripherals.
The complete system is synthesized, placed, and routed for an AMD VCU118 FPGA board~\cite{VCU118} using Vivado 2024.2 at $f_{\mathrm{clk}}=100\,\mathrm{MHz}$ with the same implementation strategy across all configurations.
\refTable{tab:fpga_resource_breakdown} shows the post-route resources of a representative $d=7$, $p_{\mathrm{phys}}=10^{-3}$ prototype integrating two Sampling Units and 20 DUTs.
The 20 DUTs account for 80.6\% of the system LUTs, while the CHEQ Evaluation Engine accounts for 15.9\%.
Within the engine, the DIU dominates LUT usage at $14.0\%$ of the \LKW{system LUTs} because each DUT requires a Decoder Bridge and Tag Queue, whereas the two Sampling Units and Mechanism Memory together use only $0.9\%$.

\begin{table}[t]
\vskip -2pt
\centering
\caption{FPGA resource breakdown of CHEQ.}
\vskip -5pt
\setlength{\tabcolsep}{4pt}
\renewcommand{\arraystretch}{1.2}
\begin{tabular}{l|rrrr}
\Xhline{0.5pt}
Component & LUTs & FFs & BRAMs & URAMs \\ \Xhline{0.4pt}

Rocket Core
  & 15,246 & 9,755 & 12.0 & 0 \\

System IPs
  & 14,329 & 20,675 & 41.5 & 0 \\

Interconnect
  & 5,316 & 8,955 & 0 & 0 \\

\Xhline{0.1pt}
\rowcolor[HTML]{EFEFEF}
CHEQ Evaluation Engine
  & 160,191 & 119,035 & 41.0 & 26 \\

\quad $\llcorner$ $2{\times}$ Sampling Unit
  & 9,032 & 5,608 & 2.0 & 24 \\

\quad $\llcorner$ Mechanism Memory
  & 42 & 121 & 0 & 2 \\

\quad $\llcorner$ Scoring Unit
  & 9,938 & 12,698 & 39.0 & 0 \\

\quad $\llcorner$ Decoder Interfacing Unit
  & 141,179 & 100,608 & 0 & 0 \\

\Xhline{0.1pt}
$20{\times}$ Helios Decoders (DUTs)
  & 810,284 & 224,682 & 0 & 0 \\

\Xhline{0.5pt}
\end{tabular}
\label{tab:fpga_resource_breakdown}
\vskip -14pt
\end{table}

\vspace{0.1em}
\noindent\textbf{Workloads.}
We use rotated surface-code memory-$Z$ circuits generated by Stim v1.16.0~\cite{Stim:Quantum21}.
Each circuit initializes $\lvert 0_L\rangle$, performs $r$ syndrome-extraction rounds, and measures $Z_L$.
We evaluate $d\in\{7,9,11,13\}$ with $r=d$ and set the four generator noise parameters---post-Clifford depolarization, pre-round data-qubit depolarization, reset errors, and measurement errors---to $p_{\mathrm{phys}}\in\{10^{-3},5\times10^{-4}\}$.
No additional idle noise is applied, and DEMs {keep hyperedges, with identical-target mechanisms merged.}
\refTable{tab:setup} summarizes the workload parameters, per-DUT resources, and service intervals; for each $d$, detector, observable, and mechanism counts are independent of $p_{\mathrm{phys}}$, with $(d^2-1)r$ detectors and $\Lambda=\sum_i p_i$ approximately proportional to $p_{\mathrm{phys}}$.

\vspace{0.1em}
\noindent\textbf{DUT configuration and timing.}
The DUT is the publicly available circuit-level Helios union-find decoder~\cite{Liyanage:TQE24}, with its decoding RTL unchanged.
Helios decodes a finite syndrome window and accepts the next window only after emitting the final correction word of the current one.
Its decoding graph is configured to match the DEM circuit and detector definitions by closing the temporal boundary preceding round 0, which is open in the released configuration but has no matching boundary before the target DEM's fixed initialization.
The correspondence between the DIU mapping and decoding graph is verified by exhaustively injecting every DEM mechanism.

The LUT and FF entries in \refTable{tab:setup} are for one standalone Helios DUT.
In a separate run that keeps the DUT continuously supplied, $C_{\mathrm{dec}}$ is measured at the DUT interface as the average interval from the start of one shot's input transfer until the DUT can accept the next shot.
It therefore includes input loading, decoding, result output, and the transition to next-shot readiness, unlike the post-input response latency $T_{\mathrm{dec}}$ in \refSection{sec:dec_eval}.
To capture $T_{\mathrm{dec}}$ for every shot, we use a cycle-indexed counter bank occupying one additional BRAM; this auxiliary measurement logic is excluded from \refTable{tab:fpga_resource_breakdown}.

\vspace{-0.3em}
\subsection{Results and Analysis}\label{sec:eval}
\vspace{-0.3em}

\begin{table}[!t]
\vskip -2pt
  \centering
  \caption{Workload and DUT characteristics.}
  \label{tab:setup}
  \vskip -7pt
  \setlength{\tabcolsep}{2pt}
  \renewcommand{\arraystretch}{1.05}
  \footnotesize
  \begin{tabular*}{1\columnwidth}{@{}l@{\hspace{3pt}}l@{\extracolsep{\fill}}rrrr@{}}
    \midrule
    \multicolumn{2}{@{}l}{Metric} & $d=7$ & $d=9$ & $d=11$ & $d=13$~ \\
    \midrule
    \multirow{5}{*}{Workload~~}
        & Detectors                         & 336    & 720     & 1,320  & 2,184~   \\
        & Observables                       & 1      & 1       & 1      & 1~       \\
        & Mechanisms                        & 5,471  & 12,705  & 24,483 & 41,909~  \\
        & $\Lambda$ ($10^{-3}$)             & 2.43   & 5.24    & 9.66  & 16.04~   \\
        & $\Lambda$ ($5\times10^{-4}$)      & 1.22 & 2.62 & 4.83 & 8.02~ \\
    \midrule
    \multirow{4}{*}{\shortstack[l]{DUT\\(Helios)}}
        & LUTs                              & 41,341 & 102,521 & 202,442 & 368,164~ \\
        & FFs                               & 11,272 & 24,680  & 45,612  & 80,400~  \\
        & $C_{\mathrm{dec}}$ ($10^{-3}$)    & 40.43  & 49.51   & 58.17  & 66.83~ \\
        & $C_{\mathrm{dec}}$ ($5\times10^{-4}$)
                                             & 37.81 & 46.11 & 53.99 & 61.67~ \\
    \midrule
  \end{tabular*}
  \vspace{2pt}
  \raggedleft
  \scriptsize
  Parentheses denote $p_{\mathrm{phys}}$; $C_{\mathrm{dec}}$ is in cycles/shot.
  \vskip -6pt
\end{table}

\begin{table}[t]
\caption{Post-synthesis Sampling Unit resources by $W$ and $d$.}
\label{tab:imp_sampling}
\vskip -4pt
\centering
\scriptsize 
\renewcommand{\arraystretch}{1.15}

\begin{tabular*}{0.98\columnwidth}{@{\extracolsep{\fill}}c|cc|cc@{}}
\Xhline{0.5pt}
~$d$ & $W$ & LUT / FF / URAM~~ & $W$ & LUT / FF / URAM~~ \\
\Xhline{0.4pt}
\multirow{3}{*}{~7} 
 & 1k  & 4,614 / 2,860 / 12~~ & 8k   & 4,660 / 2,925 / 22~~ \\
 & 2k  & 4,652 / 2,881 / 12~~ & 16k  & 4,657 / 2,944 / 42~~ \\
 & 4k  & 4,635 / 2,899 / 12~~ & 32k  & 4,680 / 2,969 / 82~~ \\
\hline
\multirow{3}{*}{~9} 
 & 1k  & 6,250 / 4,084 / 26~~ & 8k   & 6,321 / 4,132 / 48~~ \\
 & 2k  & 6,228 / 4,097 / 26~~ & 16k  & 6,310 / 4,260 / 92~~ \\
 & 4k  & 6,301 / 4,118 / 26~~ & 32k  & 6,318 / 4,168 / 180~~ \\
\hline
\multirow{3}{*}{~11} 
 & 1k  & 10,611 / 6,009 / 44~~ & 8k  & 10,713 / 6,062 / 82~~ \\
 & 2k  & 10,330 / 5,993 / 44~~ & 16k & 10,682 / 6,077 / 158~~ \\
 & 4k  & 10,699 / 6,058 / 44~~ & 32k & 10,693 / 6,090 / 310~~ \\
\hline
\multirow{3}{*}{~13} 
 & 1k  & 17,616 / 8,708 / 73~~ & 8k  & 17,649 / 8,769 / 135~~ \\
 & 2k  & 17,647 / 8,729 / 73~~ & 16k & 17,673 / 8,783 / 259~~ \\
 & 4k  & 17,835 / 8,758 / 73~~ & 32k & 17,732 / 8,840 / 507~~ \\
\Xhline{0.5pt}
\end{tabular*}
\vskip -16pt
\end{table}


\begin{figure}[t]
\vskip -4pt
\centering
\includegraphics[width=0.98\columnwidth]{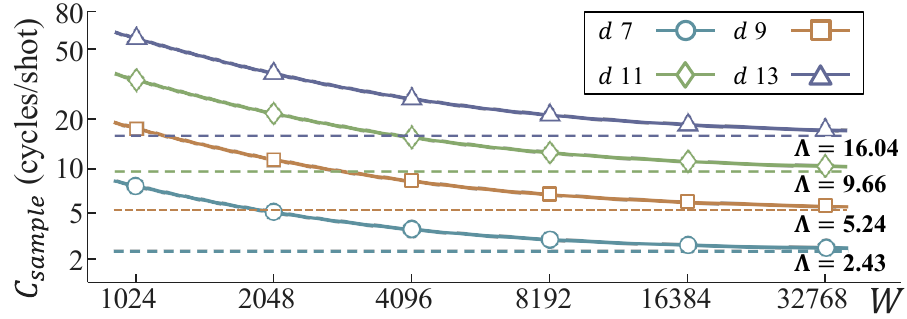}
\vskip -3pt
\caption{Sampling Unit cycles/shot versus $W$ at $p_{\mathrm{phys}}=10^{-3}$, with fits to \eqref{eq:tw} as solid lines and $\Lambda$ as dashed lines.}\label{fig:engine_eval}
\vskip -4pt
\end{figure}

\noindent\textbf{Sampling fidelity.}
We assess finite-precision fidelity at the arithmetic and RTL levels across eight workloads and six batch sizes $W\in\{2^{10},2^{11},\ldots,2^{15}\}$ (48 conditions).
Exhaustive enumeration of all $2^{32}$ Gap Generator inputs yields per-mechanism batch-average firing probabilities within $0.002\%$ of the DEM probabilities $p_i$.
Position-wise probabilities show no dependence on $W$; relative errors reach $1.49\%$ for the least probable mechanism and $0.23\%$ when firing-weighted across mechanisms.
Uniform averaging across batch positions yields the batch-average firing-rate error reported above.

RTL firing counts before Batch Memory accumulation are compared with $10^8$ Stim shots from the same DEM.
For each condition, per-mechanism two-sample $\chi^2$ statistics are summed to $M$ degrees of freedom and tested at the Bonferroni threshold $0.05/48$.
The $1.0\times10^6$--$5.0\times10^6$ RTL shots per condition provide at least $0.9$ power to detect a $1\%$ weighted RMS relative deviation $\delta$, with weights $p_i/(1-p_i)$.
No condition is rejected (minimum $p$-value $0.025$); {$\delta$ estimates reach $0.66\%$ and} Bonferroni-adjusted one-sided $95\%$ upper bounds $1.07\%$.
The \LKW{one-sided $95\%$ upper bound pooled over the 48 conditions} is $0.16\%$.
Batch firing-count variances {and lag-1 correlations are consistent with independent firing at $95\%$ confidence}; all 32 traced shots per condition match the Batch Memory contents.

\vspace{0.1em}
\noindent\textbf{Sampling resources and throughput.}
In \refTable{tab:imp_sampling}, increasing $W$ by $32\times$ changes LUT and FF usage by at most $3.7\%$ and $4.3\%$, whereas URAM grows beyond $W=4{,}096$ in Batch Memory allocation units of $D_u=4{,}096$.
This supports the $W$-dependent resource model in \refSection{sec:config}.
At $W=4{,}096$, a Sampling Unit uses $4.8$--$11.2\%$ of a Helios DUT's LUTs and $10.9$--$25.7\%$ of its FFs, with both ratios decreasing with $d$.

\begin{table}[t]
\centering

\caption{Shot-generation cycles and post-synthesis LUT usage of on-FPGA alternatives at $p_{\mathrm{phys}}=10^{-3}$.}
\label{tab:baseline_comp}
\vskip -4pt
\renewcommand{\arraystretch}{1.1} 
\begin{tabular*}{0.88\columnwidth}{@{\extracolsep{\fill}}c r r r r r@{}}
\Xhline{0.5pt}
& RISC-V sw. & \multicolumn{2}{c}{Comparator array} & \multicolumn{2}{c}{Sampling Unit} \\
\cmidrule(lr){2-2}\cmidrule(lr){3-4}\cmidrule(lr){5-6}
$d$ & Cycles & Cycles & LUTs & Cycles & LUTs \\[2pt] 
\Xhline{0.4pt}
7  & 1,297.0  & 23  & 61,735  & 3.79  & 4,635 \\
9  & 3,212.6  & 51  & 71,857  & 8.38  & 6,301 \\
11 & 7,522.5  & 97  & 86,632  & 15.68 & 10,699 \\
13 & 15,550.4 & 165 & 108,196 & 26.27 & 17,835 \\
\Xhline{0.5pt}
\end{tabular*}

\vspace{1.3em} 

\caption{Selected configurations and measured throughput; LUT (\%) denotes post-route device utilization.}
\label{tab:config}
\vskip -2pt
\renewcommand{\arraystretch}{1.25} 
\setlength{\tabcolsep}{0pt}       
\footnotesize

\begin{tabular*}{\columnwidth}{@{}c @{\extracolsep{\fill}} c c c c c c c c@{}}
\Xhline{0.5pt}
$d$ & $p_{\mathrm{phys}}$ & $C_{\mathrm{dec}}/\Lambda$ & $K_{\text{res}}$ & $N^*$ & $K^*$ & $W^*$ & LUT (\%) & Mshot/s \\
\Xhline{0.4pt}

\multirow{2}{*}{7}  & $10^{-3}$ &  16.64 & 20 & 2 & 20 & 4,096  & 85.0 & 49.16 \\
                    & $5\times10^{-4}$ & 30.99 & 20 & 1 & 20 & 8,192  & 84.4 & 52.82 \\ 
\Xhline{0.2pt}
\multirow{2}{*}{9}  & $10^{-3}$ &   9.45 & 8  & 1 & 8  & 16,384 & 78.4 & 16.14 \\
                    & $5\times10^{-4}$ &  17.60 & 8  & 1 & 8  & 4,096  & 78.3 & 17.31 \\
\Xhline{0.2pt}

\multirow{2}{*}{11} & $10^{-3}$ &   6.02 & 4  & 1 & 4  & 8,192  & 80.6 & 6.87 \\
                    & $5\times10^{-4}$ &  11.18 & 4  & 1 & 4  & 4,096  &  80.5 & 7.39 \\
\Xhline{0.2pt}

\multirow{2}{*}{13} & $10^{-3}$ &   4.17 & 2  & 1 & 2  & 4,096  & 74.9 & 2.99 \\
                    & $5\times10^{-4}$ &  7.69 & 2  & 1 & 2  & 4,096  & 74.9 & 3.24 \\
\Xhline{0.5pt}
\end{tabular*}
\vskip -15pt
\end{table}

\refFigure{fig:engine_eval} shows measured $C_{\mathrm{sample}}=C_{\mathrm{fill}}/W$ at $p_{\mathrm{phys}}=10^{-3}$.
Fitting \eqref{eq:tw} across all distances gives $a=1.0027$ and $b=1.0014$, both within $0.3\%$ of unity, with a maximum relative residual of $0.63\%$.
Thus, the measured implementation agrees with the one-cycle termination and firing costs in \refSection{sec:sampling_unit}.
Doubling $W$ halves $M/W$; costs are at most $1.161\Lambda$ and $1.082\Lambda$ for $W=16{,}384$ and $32{,}768$.

\vspace{0.1em}
\noindent\textbf{On-FPGA alternatives.}
\refTable{tab:baseline_comp} compares the Sampling Unit at $W=4{,}096$ with two on-FPGA alternatives at $p_{\mathrm{phys}}=10^{-3}$.
RISC-V software adds no logic, but generation takes 32--{233}$\times C_{\mathrm{dec}}$, or 59--297$\times$ with MMIO delivery.
A workload-matched 256-comparator array {testing} all $M$ mechanisms {per shot, mapped as in}~\cite{GarciaHerrero:Quantum26}, requires $6.1$--$6.3\times$ more cycles and $6.1$--$13.3\times$ more LUTs than the Sampling Unit.
In contrast, the Sampling Unit requires only 3.79--26.27 cycles/shot, remaining faster than one DUT at every distance.

\vspace{0.1em}
\noindent\textbf{Configuration and campaign throughput.}
{In \refTable{tab:config},} $K_{\mathrm{res}}=20$ {at} $d=7$ and $p_{\mathrm{phys}}=10^{-3}$ exceeds the single-unit supply limit $C_{\mathrm{dec}}/\Lambda=16.64$, requiring $N^*=2$; the other seven configurations use one Sampling Unit.
The two-unit configuration supplies 53.1 Mshot/s against a 20-DUT demand of 49.5 Mshot/s, and the second unit costs only $0.4\%$ of device LUTs while raising measured throughput {to 49.2 Mshot/s, beyond the single-unit asymptotic ceiling of 41.2}.
{At $d=9$, $K^*=8$ is below $C_{\mathrm{dec}}/\Lambda=9.45$, and $W^*=16{,}384$ alone lifts single-unit supply to 8.2 DUTs from 5.9 at $W=4{,}096$.}
At $d=7$ and 9, CLB occupancy reaches 94.3--98.5\% despite LUT headroom, indicating placement-limited $K_{\mathrm{res}}$.
All eight configurations satisfy $N^*/T(W^*)\ge K^*/C_{\mathrm{dec}}$, and measured throughput agrees with $K^*f_{\mathrm{clk}}/C_{\mathrm{dec}}$ within $0.62\%$, showing that generation, Mechanism Memory access, DIU dispatch, and scoring do not materially limit throughput.
At $p_{\mathrm{phys}}=5\times10^{-4}$, the reduced $C_{\mathrm{dec}}$ increases throughput at unchanged $K^*$.
For $p_{\mathrm{phys}}=10^{-3}$, a $10^{10}$-shot campaign completes in 3.4 minutes at $d=7$ and 56 minutes at $d=13$.

\begin{figure}[!t]
\centering
\vskip -4pt
\includegraphics[width=\columnwidth]{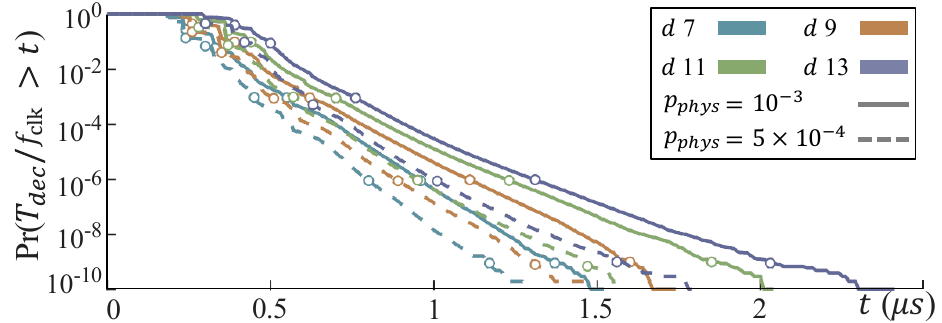}
\vskip -2pt
\caption{Fraction of $10^{10}$ shots per configuration with $T_{\mathrm{dec}}/f_{\mathrm{clk}}>t$ for $d=7,9,11,13$ and both $p_{\mathrm{phys}}$ values.}\label{fig:tail_eval}
\end{figure}


\begin{table}[t]
  \centering
  \vskip -2pt
  \caption{Logical failures and per-shot logical error rate (LER) measured over $10^{10}$ shots.}
  \label{tab:ler}
  \vskip -4pt
  \renewcommand{\arraystretch}{1.2} 
  \setlength{\tabcolsep}{0pt}       

  \begin{tabular*}{0.85\columnwidth}{@{}c @{\extracolsep{\fill}} r c r c@{}}
    \Xhline{0.5pt}
    & \multicolumn{2}{c}{$p_{\mathrm{phys}}=10^{-3}$} & \multicolumn{2}{c}{$p_{\mathrm{phys}}=5\times10^{-4}$} \\
    \cmidrule(lr){2-3}\cmidrule(l){4-5}
    $d$ & Failures & LER & Failures & LER \\[2pt] 
    \Xhline{0.4pt}
    7   & 535,171 & $5.35\times10^{-5}$ & 34,058 & $3.41\times10^{-6}$ \\
    9   & 79,408  & $7.94\times10^{-6}$ & 2,384  & $2.38\times10^{-7}$ \\
    11  & 10,627  & $1.06\times10^{-6}$ & 136    & $1.36\times10^{-8}$ \\
    13  & 1,311   & $1.31\times10^{-7}$ & 3      & $3\times10^{-10}$ \\
    \Xhline{0.5pt}
  \end{tabular*}
  \vskip -16pt
\end{table}

\vspace{0.1em}
\noindent\textbf{Latency tails and logical failures.}
\refFigure{fig:tail_eval} shows {the fraction of shots with $T_{\mathrm{dec}}/f_{\mathrm{clk}}>t$} per $(d,p_{\mathrm{phys}})$, with circles at levels $0.5$, $0.1$, $10^{-3}$, $10^{-6}$, and $10^{-9}$.
From $d=7$ to 13, the {one-round ($1\,\mu\mathrm{s}$)} violation rate rises from $4.39\times10^{-7}$ to $3.02\times10^{-5}$ at $p_{\mathrm{phys}}=10^{-3}$ and from $1.43\times10^{-8}$ to $1.02\times10^{-6}$ at $5\times10^{-4}$; {only $d=11$ and 13 at $10^{-3}$ exceed $2\,\mu\mathrm{s}$}, at rates of $3\times10^{-10}$ and $1.3\times10^{-9}$.
\refTable{tab:ler} reports LER from the same campaigns; because logical scoring is independent of deadline checks (\refTable{tab:scoring}) and no shot exceeds the 8,192-cycle maximum wait, {every shot, including late completions, receives a logical verdict}.
At $d=13$ and $p_{\mathrm{phys}}=5\times10^{-4}$, three failures give an LER of $3\times10^{-10}$ with a $95\%$ interval of $[0.6,8.8]\times10^{-10}$.
At $p_{\mathrm{phys}}=10^{-3}$, each \LKW{increase of $d$ by two} reduces LER by $6.7\times$, $7.5\times$, and $8.1\times$; from $d=7$ to 13, LER falls by $408\times$ while the $1\,\mu\mathrm{s}$ violation rate rises by $69\times$, with logical failures more frequent for $d\le9$ and deadline violations for $d\ge11$ in this DUT.
\section{Conclusion}

The proposed \textbf{CHEQ} enables concurrent hardware evaluation of QEC decoders with on-chip circuit-level noise sampling.
Geometric skipping processes sparse firing events, while batch-invariant sampling amortizes mechanism traversal and lets batch size trade memory for throughput without changing the target shot distribution.
A throughput model selects the Sampling Unit count, batch size, and decoder count under FPGA resource constraints.
Across evaluated \LKW{conditions}, the pooled one-sided $95\%$ upper bound on the weighted RMS relative firing-rate difference from Stim is $0.16\%$, while a Sampling Unit uses only $4.8\%$--$11.2\%$ of a Helios decoder's LUTs; on a VCU118, CHEQ sustains 49.2 million shots/s with 20 decoders and completes $10^{10}$ shots in 3.4 minutes.

\newpage

\bibliographystyle{IEEEtran}
\bibliography{reference}

\end{document}